%% file: template.tex
\documentclass{vgtc}                          %

\graphicspath{{figures/}{pictures/}{images/}{./}} %

\usepackage{times}                     %

\usepackage{tabu}                      %
\usepackage{booktabs}                  %
\usepackage{lipsum}                    %
\usepackage{mwe}                       %
\usepackage[T1]{fontenc}
\usepackage{cite}  %
\usepackage{tabu}                      %
\usepackage{booktabs}                  %
\usepackage{lipsum}                    %
\usepackage{mwe}                       %
\usepackage{mathptmx}                  %
\usepackage{amsmath}
\usepackage[svgnames]{xcolor}
\usepackage[framemethod=tikz]{mdframed}
\usepackage{listings}
\usepackage{xcolor}
\usepackage{mathptmx}                  %

\onlineid{0}

\vgtccategory{Research}

\vgtcinsertpkg

\input{./meta/commands.tex}

\title{Do Language Model Agents Align with Humans in Rating Visualizations? An Empirical Study}

\author{
  {\parbox{\textwidth}{\centering
    Zekai Shao\textsuperscript{1}\thanks{Co-first author. E-mail: zkshao23@m.fudan.edu.cn},
    Yi Shan\textsuperscript{1}\thanks{Co-first author. E-mail: ydan24@m.fudan.edu.cn},
    Yixuan He\textsuperscript{1},
    Yuxuan Yao\textsuperscript{1},
    Junhong Wang\textsuperscript{1}, \\
    Xiaolong (Luke) Zhang\textsuperscript{3},
    Yu Zhang\textsuperscript{2}\thanks{Corresponding author. E-mail: yuzhang94@outlook.com},
    and Siming Chen\textsuperscript{1}\thanks{Corresponding author. E-mail: simingchen@fudan.edu.cn}
  }}\\
  {\parbox{\textwidth}{\centering
    \textsuperscript{1}Fudan University, China\\
    \textsuperscript{2}University of Oxford, United Kingdom\\
    \textsuperscript{3}Pennsylvania State University, United States
  }}
}

\abstract{
    \input{./sections/0-abstract}
} %


\begin{document}
\maketitle

\input{./sections/1-introduction}
\input{./sections/2-relatedwork}

\input{./sections/3-pilotstudy}

\input{./sections/4-formalstudy}

\input{./sections/5-extratest}
\input{./sections/6-potentialscenarios}
\input{./sections/7-discussion}
\input{./sections/8-conclusion}

\bibliographystyle{abbrv-doi}

\input{./template.bbl}
% \bibliography{doc/assets/bibs/agent-eval} 

\input{./meta/appendix}
\end{document}

%% file: meta/commands.tex
\PassOptionsToPackage{svgnames}{xcolor}
\PassOptionsToPackage{table}{xcolor}
\usepackage{xcolor}
\usepackage{xspace, xpunctuate}
\usepackage{listings}  %
\usepackage{enumitem}  %
\usepackage{multirow}
\usepackage{makecell}
\definecolor{keywordcolor}{rgb}{0.56, 0.13, 0.00}
\definecolor{ndkeywordcolor}{rgb}{0.05, 0.46, 0.17}
\definecolor{commentcolor}{rgb}{0.41, 0.64, 0.70}
\definecolor{stringcolor}{rgb}{0.25, 0.44, 0.63}

\definecolor{highlightcolor}{rgb}{0.9,0.0,0.0}

\definecolor{highconfidence}{HTML}{19CAAD}
\definecolor{otherstudy}{HTML}{ECAD9E}
\definecolor{otherfield}{HTML}{E6CEAC}
\definecolor{commonsense}{HTML}{8CC7B5}
\definecolor{lowconfidence}{HTML}{E6CEAC}
\definecolor{extratest}{HTML}{D6D5B7}
\lstdefinelanguage{TypeScript}{
  keywords={typeof, new, true, false, catch, function, return, null, catch, switch, var, if, in, while, do, else, case, break, boolean},
  morekeywords={[2]{class, export, throw, implements, import, this}},
  identifierstyle=\color{black},
  sensitive=false,
  comment=[l]{//},
  morecomment=[s]{/*}{*/},
  commentstyle=\color{commentcolor}\ttfamily,
  stringstyle=\color{stringcolor}\ttfamily,
  morestring=[b]',
  morestring=[b]"
}
\definecolor{backcolour}{rgb}{0.95, 0.95, 0.92}
\definecolor{codegray}{rgb}{0.5,0.5,0.5}
\definecolor{codeblack}{rgb}{0.28,0.46,0.84}
\definecolor{codegreen}{rgb}{0,0.6,0}
\definecolor{keywordcolor}{rgb}{0.13,0.29,0.53}
\definecolor{stringcolor}{rgb}{0.31,0.60,0.02}
\definecolor{commentcolor}{rgb}{0.56,0.35,0.01}
\definecolor{ndkeywordcolor}{rgb}{0.5,0.0,0.0}
\definecolor{highlightcolor}{rgb}{0.8,0.8,0.8}

\lstdefinelanguage{Python}{
  morekeywords={def, return, True, False, None, if, else, elif, while, for, break, continue, try, except, with, as, import, from, class, pass, yield, raise, in, and, or, is, not},
  morekeywords={[2]{print, exec, lambda}},
  sensitive=true,
  morecomment=[l]{\#},
  morestring=[b]',
  morestring=[b]",
  identifierstyle=\color{black},
  commentstyle=\color{commentcolor}\ttfamily,
  stringstyle=\color{stringcolor}\ttfamily,
  keywordstyle=\color{keywordcolor}\bfseries,
  keywordstyle={[2]{\color{ndkeywordcolor}\bfseries}},
}

\setlist{noitemsep,parsep=0pt,partopsep=0pt}

\usepackage{soul}

\newcommand{\revised}[1]{{\color{black} #1}}
\newcommand{\hyx}[1]{{\color{black} #1}}

\usepackage[group-separator={$,$}]{siunitx}

%% file: sections/0-abstract.tex
Large language models encode knowledge in various domains and demonstrate the ability to understand visualizations.
They may also capture visualization design knowledge and potentially help reduce the cost of formative studies.
However, it remains a question whether large language models are capable of predicting human feedback on visualizations. 
To investigate this question, we conducted three studies to examine whether large model-based agents can simulate human ratings in visualization tasks.
The first study, replicating a published study involving human subjects, shows agents are promising in conducting human-like reasoning and rating, and its result guides the subsequent experimental design.
The second study repeated six human-subject studies reported in literature on subjective ratings, but replacing human participants with agents.
Consulting with five human experts, this study demonstrates that the alignment of agent ratings with human ratings positively correlates with the confidence levels of the experts before the experiments.
The third study tests commonly used techniques for enhancing agents, including preprocessing visual and textual inputs, and knowledge injection.
The results reveal the issues of these techniques in robustness and potential induction of biases.
The three studies indicate that language model-based agents can potentially simulate human ratings in visualization experiments, provided that they are guided by high-confidence hypotheses from expert evaluators.
Additionally, we demonstrate the usage scenario of swiftly evaluating prototypes with agents.
We discuss insights and future directions for evaluating and improving the alignment of agent ratings with human ratings. 
We note that simulation may only serve as complements and cannot replace user studies.

%% file: sections/1-introduction.tex
\section{Introduction}

Large language models (LLMs) have demonstrated impressive capabilities in encoding visual knowledge and understanding visualizations. 
They are integrated into the pipelines of visualization generation~\cite{chen_vis24_viseval} and recommendation~\cite{wang_llm4vis_2023}, and function as agents for editing visualizations~\cite{liu_ava_2023}.
A fundamental question is how effectively language models understand visualization, in terms of alignment with human interpretations.
Through answering this question, we may gain new insights into visualization perception research~\cite{yang_chi23_how_can_dnn} and may develop techniques of using language models as low-cost proxies for human subjects in certain scenarios.

Recent works have explored the alignment of language model-based agents with human interpretation in visualization tasks.
Some investigate how well model response aligns with human responses in literacy tests by chart question answering~\cite{li2024visualizationliteracy, verma2024evaluating}.
However, the alignments between agents and humans in empirical studies for usability tests or perception tasks are underexplored.
A recent study~\cite{wang_vis24_howaligned} compared LLM-generated and human-made chart takeaways on bar charts with varying layouts to examine LLMs' sensitivity to visualization design choices.
However, their study focused on sensitivity to layout as a perception task and takeaway as a response type, leaving other aspects unexplored.

In this paper, we empirically investigate LLM agents' ability to provide feedback on extensive visualization tasks.
Ratings, as a common feedback mechanism in empirical studies, are often used to reflect user perception and are prevalent in limited public empirical datasets.
Comparing the alignment between agent ratings and human responses in extensive visualization tasks may help identify agent-based evaluation schemes to complement human rating experiments.
Therefore, we design a three-stage study on simulating human ratings in visualization tasks to answer the following research questions (\textbf{RQs}).

\begin{itemize}[leftmargin=10pt]
    \item \textbf{RQ1}: What are the characteristics of agent ratings?
    \item \textbf{RQ2}: How well do agent and human ratings align?
    \item \textbf{RQ3}: How to better align agent and human ratings?
\end{itemize}

In the first study (\cref {sec: pilot study}) for \textbf{RQ1} that reproduced a human-subject experiment on time series visualization~\cite{adnan_investigating_2016} by GPT-4V~\cite{openai_gpt-4_2023} we found that the model could simulate human reasoning to give human-like rating values but failed to simulate diverse user profiles.
These findings guided the following study design.

In the second study (\cref{sec: rq2}) for \textbf{RQ2}, we carefully selected five public studies that provided open-source data on the Open Science Framework (OSF)~\cite{osf} and replicated the original experimental procedures to understand the alignment of agent ratings with human responses.
We analyzed the alignment of agent ratings with human responses in the replication of six experiments.
By also inviting five experts in visualization evaluation to code confidence for individual experiments, we suggest that the alignment of agent ratings with human responses may positively correlate with expert pre-experiment confidence.
Agents, as aggregations of historically trained knowledge, can mimic human responses but cannot replace human subjects.

In the third study (\cref{sec: rq3}) for \textbf{RQ3}, we conducted four additional experiments and observed interesting outcomes.
We found that automatic external knowledge injection may enhance the alignment of agent ratings with human ones.
However, emphasizing variables intended for comparison in prompts or presenting text-only inputs without the corresponding visualizations might lead to obvious biases from original ratings.

Additionally, we proposed an experimental scenario that involved both human-subject and agent-based studies, illustrating agents' potential in rapidly prototyping experimental designs.

Our contributions are summarized as follows:

\begin{itemize}[leftmargin=10pt]
    \item We design three studies based on existing experiments with human subjects to illustrate the relative alignment between agent ratings and human responses (\cref{sec: pilot study}), demonstrate how this alignment correlates with experts' confidence before experiments (\cref{sec: rq2}), and explore the potential for further enhancing alignment through constructive strategies (\cref{sec: rq3}).
    
    \item We demonstrate the potential scenarios for agents in swiftly evaluating prototypes (\cref{sec: potential scenarios}) and discuss future directions for simulating visualization experiments (\cref{sec: discussion}).
    
\end{itemize}

The examples of prompts and code can be found in the appendix and the complete code is available in \url{https://github.com/ZekaiShao25/Agents-Ratings-in-VIS-Experiments}.

%% file: sections/2-relatedwork.tex
\section{Related Work}

We review the related work on visualization evaluation, language models and agents, and \revised{machine understanding of visualizations}.

\subsection{Visualization Evaluation}

User evaluation is a long-standing topic in visualization research. 
\revised{Prior work categorize evaluations in terms of purpose, subject, and scenario.
Andrews~\cite{andrews_evaluation_2008} introduced a framework for categorizing evaluation methods based on purpose and subject. 
Lam et al.~\cite{lam_empirical_2012} expanded on this classification by adding seven evaluation scenarios. 
Isenberg et al.~\cite{isenberg_systematic_2013} further included another scenario.
From the perspective of collected data types~\cite{24years-eval,creswell2017research}, empirical study data includes quantitative data, such as accuracy, task completion time, and questionnaire ratings, as well as qualitative data, such as think-aloud results and interview records.
We focus our investigation on the alignment of ratings simulated by large model-based agents with human data.
}

Empirical studies in visualization research span many topics, such as dimension reduction algorithms~\cite{xia_revisiting_2021}, network representations~\cite{sarma_evaluating_2023}, time series~\cite{adnan_investigating_2016,di_bartolomeo_evaluating_2020}, parallel coordinates~\cite{johansson_evaluation_2016}, aesthetics~\cite{he_design_2024}, uncertainty~\cite{hullman_pursuit_2019, sarma_evaluating_2023}, and visualization literacy~\cite{ge_calvi_2023}.
Additionally, some work has assessed interactive visualization systems with case studies~\cite{kang_evaluating_2009,kang_how_2011} and interaction log analysis~\cite{kahng_how_2020,guo_case_2016}.
\revised{We limit the scope of our study to rate basic 2D visualizations without interaction.}

\subsection{Language Models and Agents}

LLMs, such as GPT-4 (no vision)~\cite{openai_gpt-4_2023} and Llama~\cite{touvron_llama_2023}, have shown exceptional abilities. 
LMMs, such as GPT-4V and GPT-4o, have exhibited visual understanding abilities and general intelligence~\cite{yang_dawn_2023}, employed as autonomous agents for general tasks such as web navigation~\cite{zeng_agenttuning_2023} and interpreting webpage screenshots~\cite{yang_set--mark_2023, zheng_GPT-4Vision_2024}. 
To alleviate the hallucination problem, retrieval-augmented generation (RAG)~\cite{gao_retrieval-augmented_2024} improves accuracy and relevance, particularly in fields such as scientific research~\cite{lala_paperqa_2023}. 
Additionally, crowdsourcing-inspired workflows have been proposed to optimize LLM chains for better performance~\cite{grunde-mclaughlin_designing_2023}.
These works motivated us to investigate large models in simulating human ratings and inspired the design of our agent-based procedure.

LLM is also applied to visualization generation~\cite{chen_vis24_viseval,Li2024Visualization} and recommendation~\cite{wang_llm4vis_2023}, chart summarization~\cite{ko_natural_2024,chartinsighter}, enhancing visual analytics systems~\cite{tailormind,leva}, and visual storytelling~\cite{shen_data_2024, narrativeplayer}. 
Besides integrating LLMs into automatic pipelines, Liu et al.~\cite{liu_ava_2023} utilized LLMs as autonomous agents for visualization modification.
In this work, we study the alignment between large model-based agents and human ratings.

\revised{\subsection{Machine Understanding of Visualizations}

As machine learning advances, visualization community has adopted models to enhance the understanding of visualizations. 
Previous research has employed information theory~\cite{chen_information-theoretic_2010} and visual saliency~\cite{matzen_data_2018} automatic quality evaluation or explored models like CNNs for visualization perception tasks~\cite{haehn_tvcg2019_evaluating,yang_chi23_how_can_dnn, cnn2024}.
For instance, Yang et al.~\cite{yang_chi23_how_can_dnn} demonstrated the potential of fine-tuned CNNs in predicting and generalizing human correlation judgments.

With LLMs' emergence, researchers have extended these studies to evaluate their capabilities in visualization tasks that closely align with human-subjective study settings, including low-level analytics~\cite{xu2024exploring}, literacy assessments~\cite{li2024visualizationliteracy, verma2024evaluating}, misleading visualization detection~\cite{lo_vis24_howgood}, and chart takeaway generation~\cite{wang_vis24_howaligned}.
These studies often investigate parameters and settings such as different prompts and models. 
They can be categorized into two primary approaches: one focusing on task automation, aiming to maximize LLM accuracy~\cite{xu2024exploring, lo_vis24_howgood}, and the other also measuring alignment between LLM outputs and human responses~\cite{li2024visualizationliteracy, verma2024evaluating, wang_vis24_howaligned}.
Beyond direct prompt-based models, Wang et al.~\cite{wang_vis24_dracogpt} proposed DracoGPT by using Draco to extract design preferences of LLMs, and Cui et al.\cite{cui_vis24_pitfalls} generated datasets for visualization literacy tests with LLMs.

Our work aligns with the contemporaneous research on human-LLM alignment on visualization perception tasks, but centers on the relatively underexplored area of subjective ratings as a feedback type. 
Additionally, our empirical study spans multiple public empirical studies involving subjective ratings.
We also explore the potential of RAG-enhanced agents to improve alignments and address future application scenarios.
}

%% file: sections/3-pilotstudy.tex
\section{Study I: \revised{What are the characteristics of agent ratings?}}
\label{sec: pilot study}
\label{sec: pilot study-ratings}

We first conducted Study I to replicate an experiment reported in Adnan et al.'s work published in ACM CHI 2016~\cite{adnan_investigating_2016}.
Our Study I explored the alignment between agent-simulated ratings and the ratings from human subjects.
We prioritized three factors for selecting the existing experiment:

\begin{itemize}[leftmargin=*]
    \item ease of experimental replication,
    \item categorization as 2D visualization perception tasks without interaction, and
    \item availability of collected and analyzed human ratings.
\end{itemize}

We chose to replicate the results on the improvement of user experience by a visualization design for time series data.

\revised{
\subsection{Original Study}

\textbf{Materials and Procedure.} The original study employed a within-subjects design for all independent variables:

\begin{itemize}[leftmargin=*]
    \item 4 interaction scenarios (no interaction, highlighting, tooltips, and highlighting-tooltips),
    \item 3 visual encodings (position, color, and area),
    \item 2 coordinate systems (Cartesian and Polar), and
    \item 4 tasks (maxima, minima, comparison, and trend detection).
\end{itemize}

The time series visualizations presented to the participants are created from synthetic time series datasets.
The maxima and minima tasks are to identify the highest and lowest value in the visualization, respectively.
The comparison task is to compare the aggregated value of two slices of the time series.
The trend detection task is to identify a subset of the data with the smallest upward trend.

\textbf{Original results.}
The authors analyzed the outcomes of completion time, accuracy, confidence level, and ease of use and obtained some conclusions (\textbf{C}).
We focus on their conclusions on the impact of visual encodings on ratings, which are listed as follows.
The area encoding received lower scores in other tasks (\textbf{C1.1}) but excelled in comparison tasks (\textbf{C1.2}).
}

\subsection{Agent-Based Study}

Our agent-based study excluded one independent variable, interaction, and one value of another variable, the polar coordinates, in the original design, due to the impracticality of enabling the agent to perform interactions and understand uncommon visualizations. 
We focused on having the agent work with line charts, heatmaps, and icicles, and evaluating agent-simulated ratings of ease-of-use and confidence level.
We further explored using GPT-4 to simulate rating for pixel-based visualization tasks, emphasizing the alignment with the results on ease-of-use and confidence levels by human subjects in the original study.

Our goal is to align the agent's ratings with established results (\textbf{C1.1, C1.2}) and also offer human-like reasoning in its explanations.
To achieve this goal and utilize its internal knowledge, we structured the task completion in three steps.
We prompted the agent to output its internal knowledge about the given visualization designs, which is incorporated into the input of the next step.
Then, we presented the task description and visualization images, prompting the agent to outline the steps for task completion and potential challenges. 
Next, we bundled the output from two steps and rating criteria as the final step's input, and asked the agent to assign ratings.
We experimented with one and three visualizations per input, respectively. 
We observed that ratings can be identical when given one image per session, but show clear differences when collectively given three at a time, with the latter results being more reasonable.
This observation shows that \textbf{agents differ from humans as they apply inconsistent standards for different experiment settings across sessions}.
This finding also affected our subsequent procedure design (\cref{sec: within-subjects variables}).
Thus, we inputted three visualizations together to GPT-4V for each task with 30 repeated trials.

\begin{figure}[ht]
  \centering
  \includegraphics[width=\linewidth]{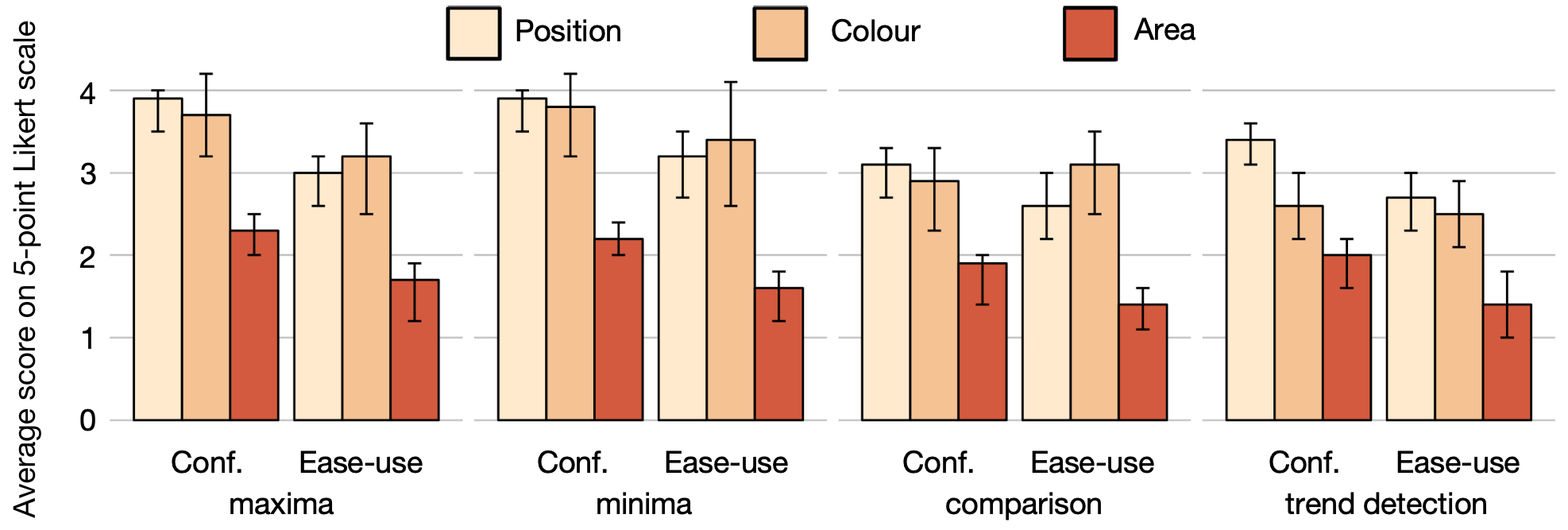}
  \caption{
    Agent ratings for time series visualization tasks. 
  }
  \label{fig:ratings}
\end{figure}

The agent's ratings, as shown in \cref{fig:ratings}, indicate that for tasks other than comparison, agents assigned significantly lower scores to area encoding, aligning with \textbf{C1.1}. 
For instance, GPT-4V describes that identifying maxima tasks is most efficiently done by locating the highest point on a line chart. 
Heatmaps rank second due to their graduated color mapping, while the icicle plot presents the greatest challenge due to the narrow width of each block and non-adjacent comparison. 
However, for comparison tasks, agents also concluded similarly, which contradicts \textbf{C1.2}.
It overemphasized aggregating daily-level blocks for weekly sales calculations, overlooking the possibility of directly comparing weekly sales between two middle tiers. 

We noted GPT-4V's knowledge in step 1 that ``for icicle charts, child nodes are added below their respective parents.'' 
This implies that the unchanged outcomes may stem from the added knowledge not clearly explaining the quantitative relationship between parent and child nodes.
Thus, we replace the self-prompted internal knowledge with ``For the icicle plot, to aggregate child nodes, consider the size of the parent node'', as the partial input of step 2.
As anticipated, this adjustment led GPT-4V to propose logical completion steps and rate the icicle plot's confidence to the highest level.
Given these insights, we \textbf{remain open to the idea of employing agents for subjective rating}.

We then aimed to assess whether the agent could embody various roles to mimic diversity within or across groups. 
Based on the original study's 24 participant descriptions, we created 24 distinct user profiles varying in gender, age, identity, and education, incorporating these user profiles into the first step's input. 
Yet, varying profiles scarcely affected the outcomes, suggesting GPT-4V's limited grasp on how different demographics influence user behavior, thus failing to represent individual differences.
Additionally, we adjusted the agent's internal parameter (temperature) to increase the diversity of responses, yet the inherent randomness of the model could not accurately simulate user diversity. 
We conclude that while agents may simulate the behaviors of an average user, \textbf{it is still challenging to produce data mimicking diverse users in visualization tasks.}

%% file: sections/4-formalstudy.tex
\section{Study II: How well do agent and human ratings align?}
\label{sec: rq2}
We conducted experiments to investigate the alignment between agent ratings and human responses align by comparing whether they lead to consistent conclusions in specific tasks (\textbf{RQ2}).

\subsection{Data Preparation}
\label{sec:data-preparation}

\hyx{

We gathered a wide array of visualization evaluation literature. 
Similar to constraints in Study I, we focused on public studies with complete original user study materials. 
We limited the scope of papers to prominent conferences and journals: VIS, TVCG, EuroVis, CGF, and CHI. The criteria and process for the inclusion of literature can be concluded as follows:

\begin{enumerate}[leftmargin=*]
    \item \textbf{Keyword matching.} We first applied keyword-based regular matching, requiring titles or abstracts to include terms like ``evaluate,'' ``investigate,'' ``user,'' ``viewer,'' ``crowdsource,'' ``assess,'' or their variations.
    \item \textbf{OSF filtering.} We then filtered the matched papers for those with experimental materials on the OSF to reduce replication costs and increase the reliability of our evaluation dataset.
    \item \textbf{Manual filtering.} Finally, we manually reviewed all filtered papers and excluded papers not related to visualization evaluation, which are surveys, those evaluating newly proposed visualizations, and those involving interaction or 3D representations.
\end{enumerate}

The results of each step are shown in (\cref{fig: data}).
In summary, we identified five suitable papers meeting our requirements: one from CHI (Timeline~\cite{di_bartolomeo_evaluating_2020}), one from VIS (Imputation for Uncertainty~\cite{sarma_evaluating_2023}), three from TVCG (Fit Estimation~\cite{reimann_visual_2021}, Texture~\cite{he_design_2024} and Magnitude Judgement~\cite{bradley_magnitude_2024}), and none from EuroVis or CGF.
The five papers form the experimental datasets.

\begin{figure}[t]
  \centering
  \includegraphics[width=\linewidth]{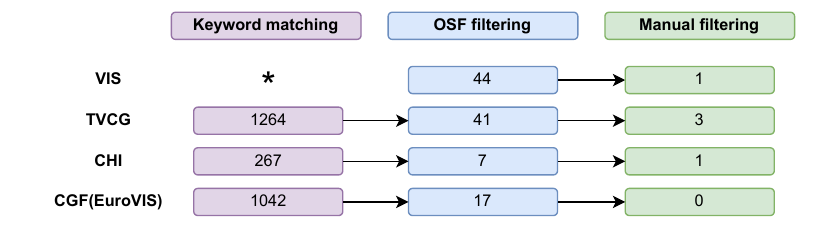}
  \caption{
    \revised{The data preparation steps.}
    The values represent the number of remaining papers after each step.
    \revised{
    For VIS papers, we skip the keyword matching step and utilized vispubdata~\cite{7583708} dataset to obtain all paper titles.}
  }
  \label{fig: data}
\end{figure}

}

Our study primarily utilized GPT-4V \revised{(temperature = 0) as the rating agent, the most powerful multimodal LLM at the time of conducting this project.
We will discuss the significance of our work for state-of-the-art LLMs in \cref{sec: discussion}.}
We also tested other LMMs across varying model sizes and types, with a goal to deepen our understanding of how model size and type influence the results.

\subsection{General Procedure}
\label{sec: general modifications}

We identified the variables involved in each experiment, such as visualizations, tasks, and data. 
We focused on simulating conclusions related to rating, which include participants' confidence level, aesthetics, and readability topics.

\subsubsection{Replicating Original Study with Prompting Agents}

To ensure the integrity of our experiments, we strive to replicate the original experiment procedure based on the open-source data as closely as possible within the constraints presented by the current capabilities of the agent. 
This replication involves inputting training tutorials and task descriptions into the prompt to provide a contextual foundation for agents' operations.
If rating is the task itself, the agent assigns points in a single operation. 
If the original study required completing the task first before ratings, then the agent proceeds in a two-operation chain: first outlining steps to complete the task, then assigning ratings based on those steps.
However, we had to modify the experimental setup in two aspects.

\label{sec: within-subjects variables}
\textbf{Transitions from between-subjects to within-subjects variables.} 
The necessity for such modifications arises from the lack of standardized rating criteria for comparing visualizations.
They result in the agent's inability to maintain a consistent standard across time and images, as humans do (see observations in \cref{sec: pilot study-ratings}).
As a result of our modification towards within-subjects variables, different sessions shared an identical experiment setting, making the results meaningful.

\textbf{Transitions from global comparison to local comparison.}
\label{sec: local comparison}
The agent's token limit restricts image input, with GPT-4V's API often failing beyond nine normal-resolution images. 
Furthermore, it's not feasible to fix a specific image as a standard and use its rating as a benchmark across multiple sessions for comparison (see observations in \cref{sec: pilot study-ratings}). 
Hence we had to limit image batches by selecting and inputting the most relevant images.

\textbf{Potential Limitations for procedures of our agent-based study.}
We shuffled the input images within one session to eliminate the learning or fatigue effect brought by our modification to within-subjects design, a practice frequently adopted in human-subject studies.
However, due to the input limit, shuffling a small set of images may not mitigate the effect as effectively as a larger set does in human studies, since images are not as disordered as in a larger shuffled set. 
Moreover, our sample size was insufficient to form a nearly uniform distribution of permutations to counterbalance the order effect.
For experiments where conclusions can be drawn by comparing images within the input limit, we attempt to replicate their findings but not all experimental phenomena. 
Conversely, current agent-based studies are not feasible for replication if the quantity of inputted images is insufficient for the conclusions.
We acknowledge these limitations above and specifically discuss how potential flexible design can alleviate them in \cref{sec: potential directions}.

\revised{
\subsubsection{Results Analysis by Comparing Alignments of Conclusion}
\textbf{Generally, we focused on the alignment between the agent's feedback and original conclusions (\textbf{C}) and hypotheses (\textbf{H}) (predefined or not),  rather than data distributions.}
Our observations show that agents' baseline of rating differs from human evaluators. 
This is especially evident when agents rate on broader scales such as 0-100, where their average ratings may endogenously higher or lower than human participants' ratings, and the differences should not be ignored.
For example, in Fit Estimation study~\cite{reimann_visual_2021}, the ratings of human raters are mainly distributed between 45-90 in the 0-100 range, while the ratings of agent raters are mainly distributed between 70-85 (\cref{fig: fit}).
In our analyses, the problem can be resolved by comparing conclusions, which avoids referring to absolute values.
In addition, our analyses were based only on sample information, where we employ empirical distribution.

}
\subsection{Experiment I: Fit Estimation}
\label{sec: fit}

\subsubsection{Original Study} 

\textbf{Materials and Procedure.} 
Reimann et al.~\cite{reimann_visual_2021} aimed to understand the patterns in visually estimating the fit between models and data. 
Two main experiments were conducted involving 62 German students to rate how well the data points fit the shown line.

The first experiment delved into the relationship between the noise level from the model and its visual perception of fit ratings.
It involved participants using an online interface to sequentially rate the fit for 30 scatterplots, presented in random order, based on a 5 (noise levels) x 6 (parallel versions with random values) within the subject. 
Using a slider bar, participants rated the fit on a scale from 0 (very low fit) to 100 (very high fit), with instructions and visualization provided. 
Each generated scatterplot featured a consistent model line and varied noise levels were achieved through adjusted y-values for each x-value.
We identified with two hypotheses.
The first is a significant influence of noise on visual estimations (\textbf{H2.1}). 
The second is a negatively accelerated relationship between noise levels (statistical deviations) and visual perception (\textbf{H2.2}), derived from Fechner’s law of psychophysics.

The second experiment explored the influence of the decentering of data points from the model on fit estimation. 
It replicated the first one's methodology, presenting participants with randomized 42 scatterplots based on a 6 (noise levels) x 7 (decentering conditions) within-subjects design. 
Each scatterplot was manipulated with noise by varying the standard deviation (sd) from 0.2 to 1.2 in steps of 0.2. 
Decentering was adjusted by changing the mean y-value with a certain ratio for each x-value, with seven conditions ranging from neutral (ratio $= 0$) to three increasing levels (ratio $= 0.25, 0.5, 0.75$) of decentering both upwards and downwards the model line. 
Evaluators expected that increased decentering would correlate with lower fit estimations (\textbf{H2.3}) at constant noise levels.
However, they were uncertain if participants would strongly weigh decentering, leading to no predefined hypothesis.

\textbf{Results.}
For the first experiment, a repeated measures ANOVA confirms that noise significantly affects visual fit estimation $(F(4, 244) = 273.04, p < .001, \eta^{2}_{p} = .82)$, with all pairwise comparisons significant (Bonferroni corrected) (\textbf{C2.1}).
Visual deviation ratings increase more sharply at lower noise levels compared to higher ones, with $t$-test confirming these differences in rate changes as significant (\textbf{C2.2}), $t(61) = 7.61, p < .001, d = 0.97$. 
These conclusions from the user study align with both two hypotheses. 

For the second experiment, an ANOVA analysis combined upward and downward decentering, shows significant effects of both decentering, $F (1.43, 87.23) = 28.25, p < .001, \eta_{p}^{2} = .32$ (\textbf{C2.3}), and noise on fit estimations without a significant interaction between these factors.
Thus, \textbf{C2.3} is fully aligned with \textbf{H2.3}.
Decentering has a robust but surprisingly small impact compared to noise (\textbf{C2.4}), as the largest difference in fit estimations ($M = 36.27, SD = 14.19$) between the highest and lowest noise levels, is significantly greater than the difference between centered and strongest decentered conditions ($M = 8.48, SD = 11.39$), $t(61) = 11.32, p < .001, d = 2.16$.

\subsubsection{Agent-based Study}
\label{sec: fit replication}
\textbf{Material and Procedure.}
For the first experiment, we verified \textbf{C2.1} and \textbf{C2.2} through 6 sets of parallel versions, where GPT-4V evaluated 5 scatterplots at varied noise levels, repeated 5 times to minimize randomness.
For the visual input, all scatterplots were generated using the open-source code that generates stimuli in OSF.
The order of the inputted five scatterplots was random, which is consistent with the original study.
For the textual prompt, we began with the role-play instruction ``You are an average user'' and concluded with ``Please list the ratings and reasons in the order of the images.''
The remaining prompts were directly sourced from the tutorial and task descriptions in the original paper and OSF.

The original second experiment further explored \textbf{C2.3} and \textbf{C2.4} totaling 42 images, which exceed the input limits as stated in \cref{sec: local comparison}.
Therefore, we performed a local comparison with three methods: comparison of decentering across four levels against a fixed noise level, comparison of decentering effects at noise levels of 0.2 and 1.2, and comparison in specific conditions (e.g., strong decentering at noise=0.2 vs. neutral decentering at noise=0.4).
The first comparison aimed to simulate \textbf{C2.3}, the second one targeted \textbf{C2.3} and \textbf{C2.4} verification, and the last emerged from the close ratings of pairs in the original experiment. 
In the first comparison, the agent rated 7 scatterplots with varying decentering but fixed noise levels, repeating 5 times.
In the second comparison, the agent rated 8 scatterplots each time, including 2 with no decentering and another random (upwards or downwards) 6 with different decentering for 50 parallel trials.
In the third comparison, the agent rated 3 scatterplots from each condition, repeating 5 times.
The prompts were the same as in the first experiment.

\begin{figure}[t]
  \centering
  \includegraphics[width=\linewidth]{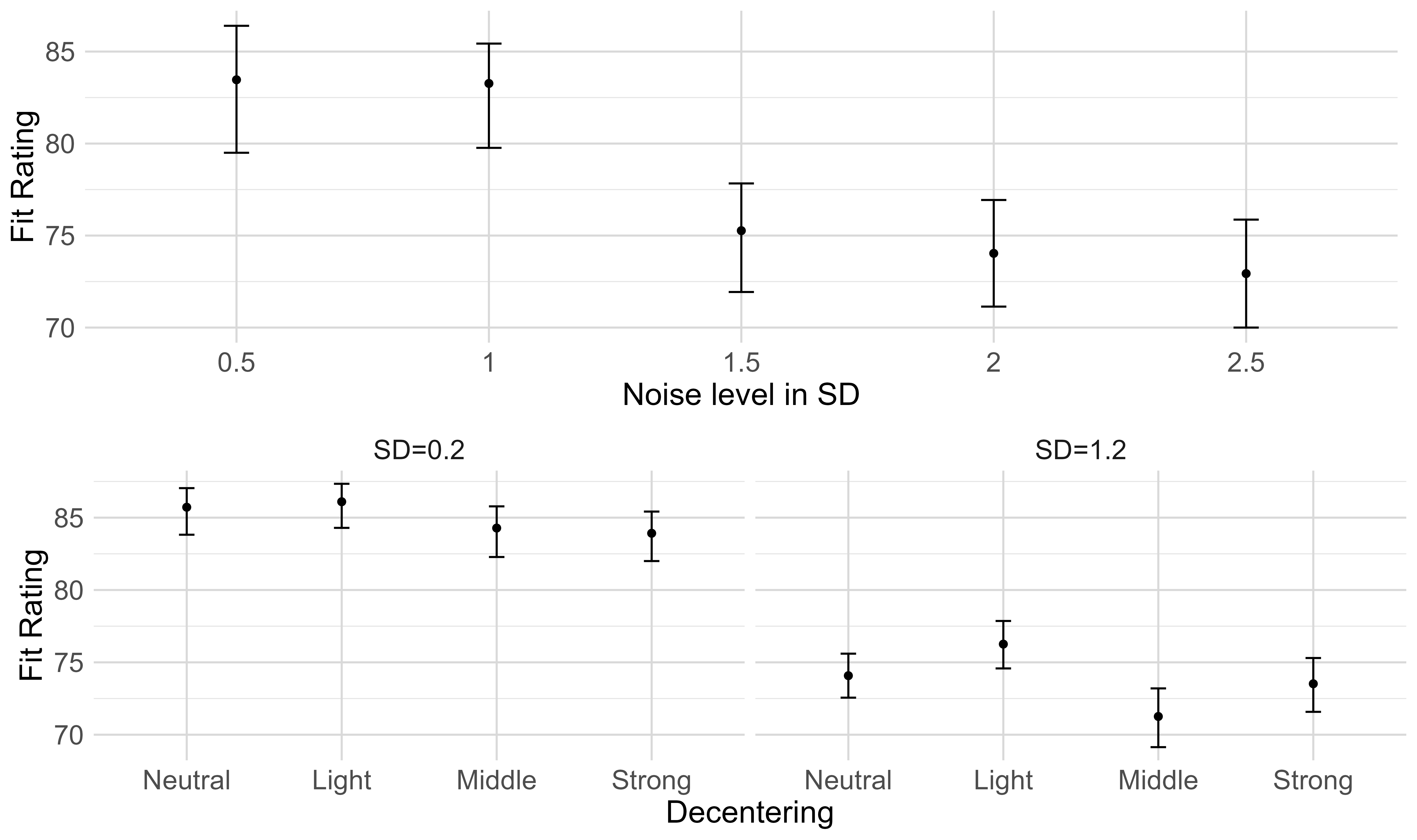}
  \caption{Results of agent ratings for fit estimations under various noise levels (top) and decentering (bottom). Error bars represent 95\% bootstrap CIs.}
  \label{fig: fit}
  \vspace{-8pt}
\end{figure}

\textbf{Results.}
As we decided to make no assumptions on the agent's output distribution across sessions, we chose not to use ANOVA or $t$-tests.
For each condition, we employed bootstrap (10,000 replicates) to estimate the 95\% CI of the means.
The results of the first agent-based experiment are shown at the top of \cref{fig: fit}. 
Agent ratings largely confirm \textbf{C2.1}, but do not support \textbf{C2.2}. 
This is because the scores show minimal change from noise levels 0.5 to 1, and experience a significant drop from 1 to 1.5, with the non-overlapping CIs indicating a significant difference between these means.

Agent-based ratings for the second comparison in the second experiment are presented at the bottom of \cref{fig: fit}. It suggests that the agent ratings support \textbf{C2.4}, as the influence of noise substantially outweighs that of decentering. 
The results partially support \textbf{C2.3}, showing that at a constant noise, strong decentering results in lower ratings compared to no decentering.
However, fit ratings do not consistently drop with increasing decentering, suggesting the agent perceives decentering less acutely than humans.
The results of other comparisons partially align with the user study, which can be found in our codes repository.
Additionally, the score with sd=1.2, displayed at the bottom of \cref{fig: fit}, is significantly lower than the score for sd=1 shown above, again showing that the agent's analytical perspective is not consistent in different experiment settings (\cref{sec: pilot study-ratings}). 
This result indicates the necessity of our modification from between-subjects to within-subjects variables.

\subsection{Experiment II: Imputation for Uncertainty}
\label{sec: uncertainty}

\subsubsection{Original Study}
\textbf{Materials and Procedures.} 
Sarma et al.~\cite{sarma_evaluating_2023} explored how different representations and uncertainty affect chart interpretation.
They used a mixed design with six representation types (no imputation, mean, CI, density, gradient, and HOPs with animation) as between-subjects treatments, and two tasks (estimating average and trend) and two proportion levels as within-subjects treatments.
The tutorial explained the representation, missing values, and their proportion. 
It stated that missing values were estimated using an appropriate imputation method, noting the inherent uncertainty of these estimates. 
Participants also performed a practice task with feedback.
Regards ratings-related results, participants’ confidence was gathered using a 5-item Likert-style question on the interface.

\textbf{Results.}
They applied a Bayesian hierarchical ordinal regression model to estimate participants' response probabilities on Likert items, analyzing the effects of representation, proportion, trial\_id, block\_id, and participant\_id.
Trial and block items were used to model learning effects, while they were marginalized as no meaningful effects were found.
The last item was used to model individual characteristics.
For ratings, they computed the likelihood of an average participant rating their confidence as 3 or higher on the Likert scale, $Pr(\geq3)$, across representation conditions.
Results showed that participants had the highest confidence with no imputation in average estimation $Pr(\geq3)=64\%$, slightly less with mean imputation $Pr(\geq3)=52\%$, and even lower in conditions showing uncertainty (\textbf{C3.1}).
In addition, there were no significant differences in confidence in trend estimation (\textbf{C3.2}).
They made the hypothesis that presenting mean imputation estimates alone would result in the highest participant confidence, regardless of task types (\textbf{H3.1, H3.2}). 
They expected that no imputations would reduce confidence due to a lack of information, with participants aware of this fact. 
They expected that explicitly showing uncertainty would diminish confidence due to the direct acknowledgment of uncertainty.
Thus, \textbf{H3.1} partially aligns with \textbf{C3.1}, while \textbf{H3.2} is disproven.

\subsubsection{Agent-based Study}
\textbf{Material and Procedure.}
Following \cref{sec: within-subjects variables}, we shifted to a within-subjects design for variables of representation.
We excluded HOP as it contains animation and the agent couldn't process it.
For one task (estimating mean or trend), the agent completed 2 (proportion) $\times$ 2 (error) $\times$ 2 (condition) $=$ 8 trials for the inputted five representations.
Each trial was repeated 5 times to reduce randomness, resulting in 40 sessions. 
For visual input, we used original images from the OSF. 
The textual prompt comprised direct copies of descriptions from the tutorial and interface, with the added role-play instruction, ``You are an average user.''
Following the procedure where participants first complete the task and then provide ratings, the agent first generated steps based on the task description and visual images, subsequently, rating based on these steps.

\textbf{Results.}
In our initial model selection process, we supposed the original regression model was impractical due to the lack of prior knowledge about the distribution or assumptions of the agent's output and other potential factors.
We discarded variables like participant and learning factors due to our modifications to the experiment procedure and tried adding random intercepts for each session.
However, the regression model did not capture data well, as posteriors were largely different from the empirical distribution.
Therefore, we employed descriptive statistics to analyze the frequency distribution of agent ratings for each imputation method, by comparing the proportion of agent ratings $\geq 3$.
\cref{fig: uncertainty}A reveals that in average estimation, the agent rated mean imputation highest ($P(\geq3)=57\%$), with baseline ($P(\geq3)=48\%$) and gradient ($P(\geq3)=50\%$) methods scoring lower, aligning with \textbf{H2.1} but not with \textbf{C2.1}. 
In trend estimation, mean, baseline, and CI received high scores with the same proportion, against both \textbf{C3.2} and \textbf{H3.2} as other methods scored significantly lower.

\subsection{Summary of Other Replication Experiments}
\label{sec: summary of other replication experiments}
We carried out four additional replication experiments on GPT-4V.
Among 9 conclusions in these experiments, the agent's ratings were only aligned with \textbf{C4.1} in Timeline (\cref{tab: summary-of-experiments}). 
We also conducted tests on open-sourced models including LLaVA-series model~\cite{liu_improved_2023} (7B and 13B) and BakLLaVA 1 (7B). 
Due to the space limit, we briefly summarize our findings here.
Detailed study design and results are available in the appendix.
\begin{itemize}[leftmargin=10pt]
    \item \textbf{Agent ratings may align with general human preferences, though it falls short of simulating detailed findings}. In Timeline~\cite{di_bartolomeo_evaluating_2020} experiment, the agent consistently preferred horizontal lines for any dataset or task type as the most readable timeline visualization.
    \item \textbf{Agent may not simulate aesthetic and textural readability.} In Texture~\cite{he_design_2024} experiment, agent ratings cannot align with any original conclusion related to texture.
    \item \textbf{Agent may focus on textual elements within images.} In Magnitude Judgement~\cite{bradley_magnitude_2024} experiment, the agent always offered an anti-human interpretation of axis manipulation by focusing on the text labels. 
    \item Open-sourced multimodal agents demonstrate poor comprehension in visualization tasks.
\end{itemize}

\subsection{Confidence Coding and Implications}
\label{sec: confidence Coding and implications}

\textbf{Confidence Coding.}
Agent ratings usually align with human data for basic visualization tasks that seem predictable (\textbf{C1.1}, \textbf{C2.1}) and often resemble expert hypotheses (\textbf{H3.1}).
This inspires us to consider the relationship between their alignment and the confidence level of human experts in forming hypotheses. 
Since the original study's expert evaluators typically did not emphasize hypothesis confidence, we consulted five external visualization experts for confidence coding.
All experts have had over three years of experience in visualization research, designed and conducted formal user studies, and had more than three publications in TVCG and VIS.
We engaged in one-on-one discussions with each expert for 40 minutes, presenting the original experiment's materials, procedures, and background information mentioned in the study (such as relevant research findings or theories from other fields), but excluding the original evaluators' hypotheses. 
We asked them to decide on forming their hypotheses for each experiment and give confidence in three-scale Likert ratings. 
If they could, they would assign a high (3) or medium (2) confidence level, indicating strong or moderate certainty in their predictions about general user performance, respectively. 
If they were unable to form a hypothesis, they assigned low (1) confidence.

For cases with hypotheses in the original study, we found that the hypotheses from five experts largely matched. 
For the few opposing hypotheses (no more than one in five), we lowered their confidence level to low.
In cases without original hypotheses, external experts' hypotheses were more diverse. 
We adjusted the confidence levels to low for experts who proposed opposing hypotheses.
This change highlights that experts do not have unified assumptions.
We averaged the modified confidence levels of the five experts for each conclusion, rounding up to determine human expert confidence, as illustrated in the ``Confidence'' column of \cref{tab: summary-of-experiments}.

\textbf{Implications.}
The results from our agent-based replication experiments (black rows in \cref{tab: summary-of-experiments}) can be summarized as follows:

\begin{itemize}[leftmargin=10pt]
    \item \textbf{Ratings from the agent often align with human results for basic tasks where experts have high confidence.} 
    In experiments where external experts had high confidence, the match score of agent ratings is $3/5$ (\textbf{C1.1}, \textbf{C2.1}, and \textbf{C4.1}); in other cases, the match score is $1/12$. 
    These three \textbf{C}s represent the most basic and general findings of each experiment, involving only a limited number of variables.
    \item \textbf{The alignment of agent ratings drops when expert confidence is moderate and relies on additional knowledge or unreliable common sense.} 
    In such cases, agent feedback never fully aligns with the conclusions, often showing partial alignment or complete opposition, regardless of original hypothesis-conclusion alignment.
    \item \textbf{Agent ratings are almost useless when experts have low confidence and set no consistent predefined hypothesis.}
    The rest are low-confidence scenarios usually without original hypotheses, where agent ratings seldom support the original conclusions, except for alignment with \textbf{C3.4}.
\end{itemize}

Our response to \textbf{RQ2} suggests that the alignment of agent ratings with human data might correlate with expert confidence for hypotheses.
Agents extract basic information from images, integrate it with task-specific text prompts, and leverage their internal historical knowledge derived from training to approximate human behavior and predict user study outcomes.

\input{./assets/tables/2-formal-study/overview-of-experiments}

%% file: assets/tables/2-formal-study/overview-of-experiments.tex
\begin{table}[t]
\centering
\caption{
A summary of our experiments on ratings investigating the alignment among hypothesis (H), conclusion (C) and agents' feedback (A), with "P" denoting a partial alignment.
Black rows represent replicate experiments for \textbf{RQ1} (\cref{sec: pilot study-ratings}) and \textbf{RQ2} (\cref{sec: rq2}), \textcolor{extratest}{light rows} represent additional experiments for \textbf{RQ3} (\cref{sec: rq3}). 
% \yz{The light color is hardly visible.}
}
\label{tab: summary-of-experiments}
\scalebox{0.75}{
\begin{tabular}{ccccccc}
\toprule
        \multirow{2}{*}{Experiment} & \multirow{2}{*}{Conclusion} & \multirow{2}{*}{Confidence} & \multirow{2}{*}{Hypothesis} & \multicolumn{3}{c}{Alignment} \\
        & & &  & H-C & H-A & C-A \\ 
\midrule
    \multirow{4}{*}{\makecell[c]{time series\\~\cite{adnan_investigating_2016}}} &  \textbf{C1.1} & H (2.93)& N &  &  & Y \\ 
     & \color{extratest}\textbf{C1.1} & \color{extratest}H \color{extratest}(2.93)& \color{extratest}N & \color{extratest} & \color{extratest} & \color{extratest}Y \\
     & \textbf{C1.2} & H (2.80) & N &  &  & N \\ 
     & \color{extratest}\textbf{C1.2} & \color{extratest}H \color{extratest}(2.80) & \color{extratest}N & \color{extratest} & \color{extratest} & \color{extratest}N \\ 
\midrule
    \multirow{4}{*}{\makecell[c]{fit \\ estimation\\~\cite{reimann_visual_2021}}} & \textbf{C2.1} & H (2.60) & Y & Y & Y & Y \\ 
     & \textbf{C2.2} & M (2.20)& Y & Y & N & N \\ 
     & \textbf{C2.3} & M (2.20) & Y & Y & P & P \\  
     & \textbf{C2.4} & L (1.40) & N &  &  & Y \\
\midrule
    \multirow{3}{*}{\makecell[c]{imputation for \\ uncertainty \\~\cite{sarma_evaluating_2023}}} & \textbf{C3.1} & M (1.80) & Y & P & Y & P \\ 
     & \color{extratest}\textbf{C3.1} & \color{extratest}M \color{extratest}(1.80)& \color{extratest}Y & \color{extratest}P & \color{extratest}P & \color{extratest}Y \\  
     & \textbf{C3.2} & L (1.40) & N & N & N & N \\ 
\midrule
    \multirow{4}{*}{timeline~\cite{di_bartolomeo_evaluating_2020}} & \textbf{C4.1} & H (3.00) & Y & Y & Y & Y \\ 
     & \textbf{C4.2} & M (2.47) & Y & N & N & P \\ 
     & \color{extratest}\textbf{C4.2} & \color{extratest}M \color{extratest}(2.47) & \color{extratest}Y & \color{extratest}N & \color{extratest}Y & \color{extratest}N \\ 
     & \textbf{C4.3} & L (1.42) & N &  &  & N \\
\midrule
    \multirow{4}{*}{texture~\cite{he_design_2024}} & \textbf{C5.1} & L (1.40) & N &  &  & N \\ 
     & \textbf{C5.2} & L (1.20) & N &  &  & N \\ 
     & \textbf{C5.3} & M (1.80) & Y & P & N & P \\ 
     & \textbf{C5.4} & M (1.90) & Y & P & N & N \\ 
\midrule
    \multirow{2}{*}{\makecell[c]{magnitude \\ judgement~\cite{bradley_magnitude_2024}}} & \textbf{C6.1} & H (2.80) & Y & Y & N & N \\ 
     & \textbf{C6.2} & M (1.60) & N &  &  & N \\
\bottomrule
\end{tabular}
}
% \vspace{-10px}
\end{table}

%% file: sections/5-extratest.tex
\section{Study III: How to better align agent and human ratings?}
\label{sec: rq3}
Leveraging the insights gained from simulating the original procedures, we also dedicated our engineering efforts to achieving more aligned results by proposing specific strategies (\textbf{RQ3}).

\subsection{How Do Visual and Textual Cues Influence Reasoning?}
\label{sec: extra test}

Previous findings highlight the agent's preference for textual cues over visual ones, with prior research showing its semantic reasoning significantly outperforms visual reasoning~\cite{fan_nphardeval4v_2024}. 
This motivated us to explore agent-based studies with minimal reliance on images. 
We began with Time Series to assess the impact of image modifications and further tested Imputation for Uncertainty to examine the role of keywords in text prompts for reasoning.

\textbf{Time Series: agent ratings rely on general perception and knowledge, but ignore visual details such as color.}
Color plays a crucial role in human perception for visualization. 
Although the agent provided human-like ratings (\cref{sec: pilot study-ratings}), its explanations lacked specific references to the color encoding of heatmap.
Consequently, we altered the visual input to grayscale images. 
The change yielded similar scoring distributions, with explanations omitting mentions of ``grayscale'' or color-related words.

\begin{figure}[t]
  \centering
  \includegraphics[width=\linewidth]{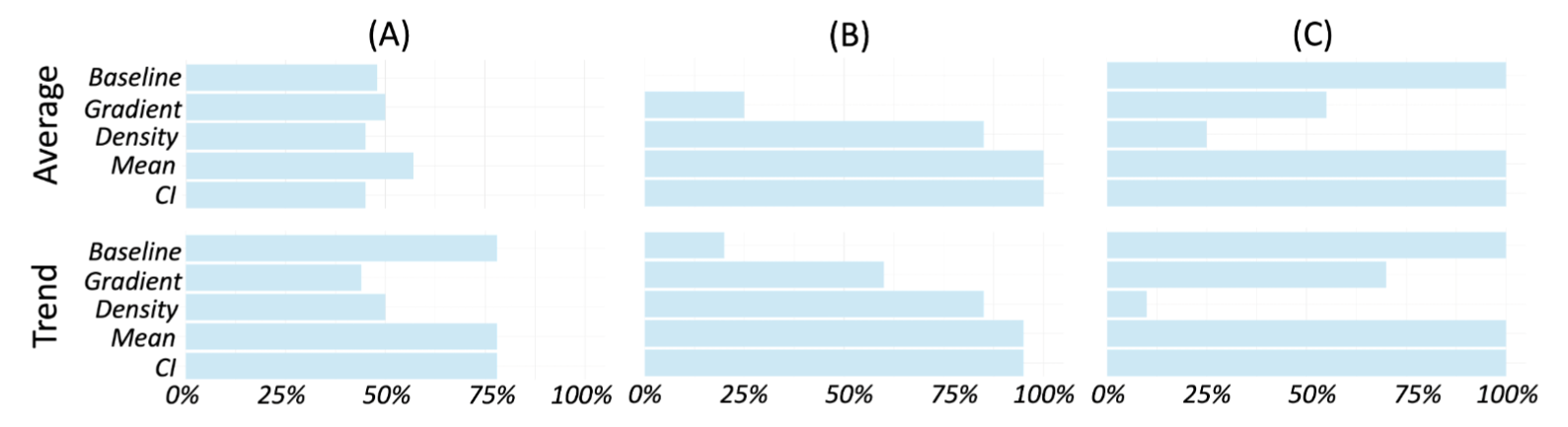}
  \caption{Proportion of agent ratings $\geq$ 3 for imputation methods addressing uncertainty. They feature two tasks (estimating the average and trend) across three conditions: (A) text and images combined, (B) text only with the same prompt, and (C) text only with a general description.}
  \label{fig: uncertainty}
  \vspace{-8px}
\end{figure}

\textbf{Imputation for Uncertainty: using only text prompts for general tasks without specific scenarios increases the risk of data pollution.}
We discarded the image inputs and organized the inputs in two forms: (1) retaining the original textual prompt, and (2) changing the detailed task and scene description in the prompt to a general description.
\cref{fig: uncertainty} shows that with images removed, the agent assigned more extreme but consistent scores, either above or below 3.
Yet, keeping the original textual cues in the prompt resulted in entirely inaccurate scores, like all baseline scores falling below 3 (``Mean'' tasks of \cref{fig: uncertainty}B). 
This occurred because textual cues carry significant detail-related information (e.g., groups in scatterplots, color), and the absence of images leads to incomplete inferences by the agent.
Conversely, directly inquiring on the study's final concerns yielded the results that are consistent with \textbf{C3.1}, with the average score for the baseline surpassing the mean (3.85 vs 3.6).
The agent noted in its explanation that ``\textit{baseline will make users feel more reliable, although this may be overconfident},'' which aligns with the original study's post-experiment analysis. This result raised a concern with data pollution in the discussion.

We also conducted experiments on Timeline. The procedures and results are included in the appendix. We found that explicit comparisons of variables may make the agent rely on the text to offer stereotypes similar to those of experts.

\subsection{Does Knowledge Injection Improve Alignment?}
\label{sec: extra engineering}

\begin{figure}[ht]
  \centering
  \includegraphics[width=0.85\linewidth]{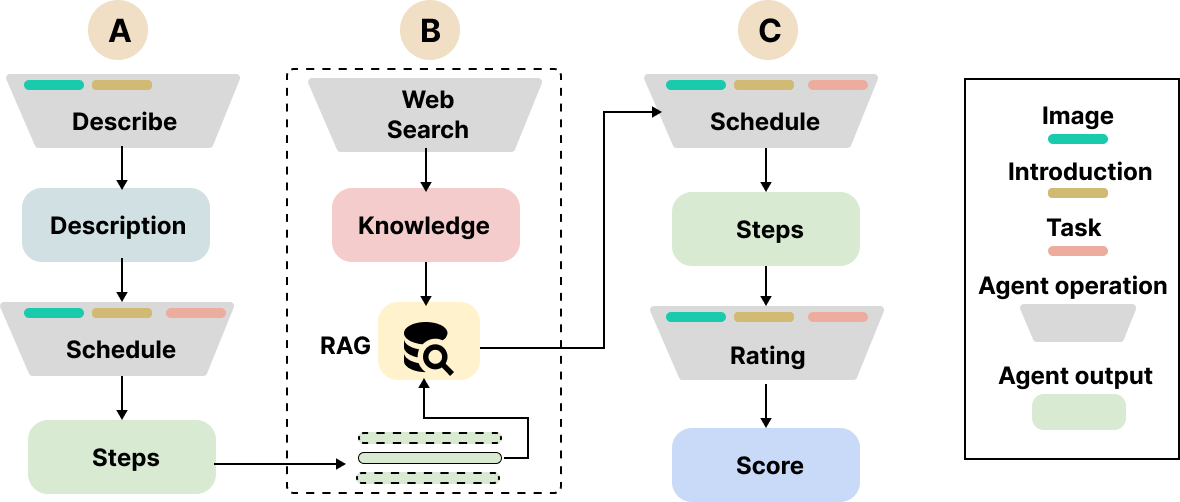}
  \caption{RAG process designed for knowledge injection in the experiment of Time Series in Study III.}
  \label{fig: rag}
  \vspace{-8px}
\end{figure}

\textbf{Experimental Design.} The success of manually adding specific knowledge to icicle plots (\cref{sec: pilot study-ratings}) and the popularity of RAG technology~\cite{gao_retrieval-augmented_2024} underscores the potential of automatic knowledge injection in our context. 
RAG technology involves three steps: 1) compiling a corpus and training a vector database; 2) querying the needed knowledge with natural language; and 3) incorporating the related knowledge into the prompt for answering.
Given the absence of a specific corpus for visualization evaluation, we employed web retrieval as our knowledge source.
The workflow of automatic knowledge-assisted answering is illustrated in \cref{fig: rag}.
The initial two steps (\cref{fig: rag}A) were the same as those in \cref{sec: pilot study-ratings}, where the agent first described visual graphs and then scheduled steps for task completion.
Different from the initial replication experiment, we inputted specific visualization names, like ``icicle plot,'' into another web-search-enhanced agent (GPT-4) and prompted it to gather related knowledge from web pages, stored in a database.
Then we split the agent-simulated steps into sentences, each serving as a query into the web-search-sourced database to retrieve knowledge blocks by semantic similarity of vector embeddings (\cref{fig: rag}B).
For each sentence query, we repeated 3 trials to identify the most related knowledge block. 
This top-1 block was then embedded in the schedule prompt to guide subsequent step generation and ensure knowledge-informed ratings (\cref{fig: rag}C).

\textbf{Experiment Results.} Incorporating three visual graphs as the inputs, our RAG-enhanced procedure yielded promising results.
While the rating distribution for other tasks remained largely the same, icicle plots in the comparison task saw significant improvement, making their overall ratings closer to the other two visual encodings (compared to \cref{fig:ratings}). 
A detailed analysis showed that although icicle plots never ranked first in the original procedure (\cref{sec: pilot study-ratings}), they occasionally achieved the highest scores in RAG-enhanced procedure (19.0\% for confidence level; 20.7\% for ease of use).
This improvement can be attributed to the action to correct a previously incorrect step involving icicle plot comparisons. 
This approach suggests ``\textit{aggregating all sizes of the weeks},'' while the relevant search found ``\textit{where the size of parent nodes directly corresponds to the sum of their children's sizes.}.''
This knowledge guidance effectively alters the steps and ratings consistently.

\textbf{Implications.} This result illustrates the potential help of using external knowledge injection.
Yet, we recognize potential issues with this approach. 
The accuracy and relevance of web search information to the agent's erroneous steps are uncertain, limiting the generalizability of our method. 
Furthermore, even with improvements from automatic external knowledge guidance, the results seem to need more scrutiny and verification before being used. 

%% file: sections/6-potentialscenarios.tex
\section{Potential Scenarios: Fast Prototyping Study}
\label{sec: potential scenarios}

\begin{figure}[ht]
  \centering
  \includegraphics[width=\linewidth]{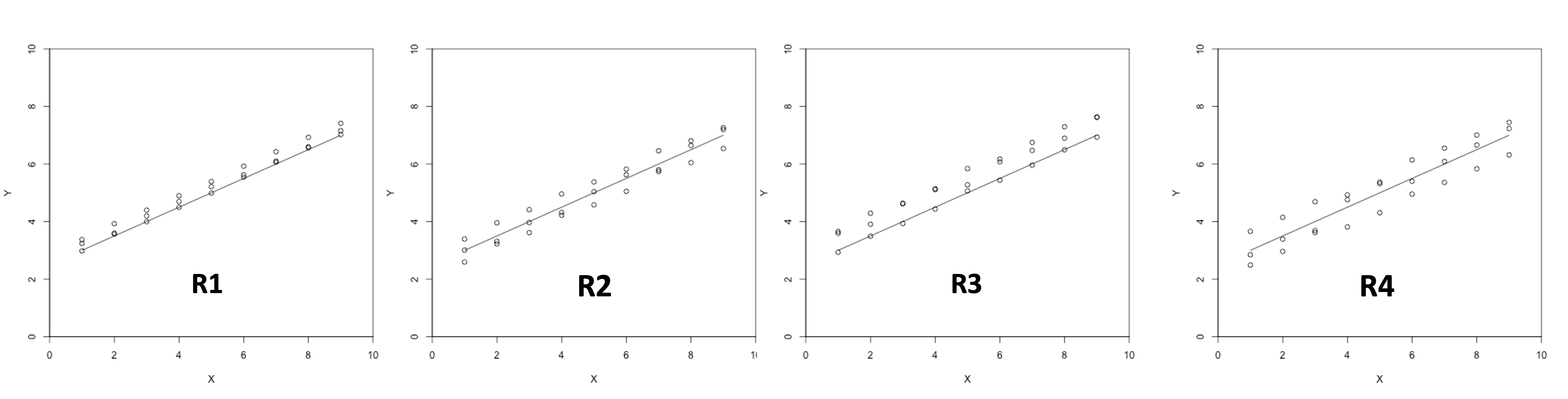}
  \caption{
    Scatterplots: strong decentering (0.99) with 0.2 noise (R1) and 0.4 noise (R3); no decentering with 0.4 noise (R2) and 0.6 noise (R4).
  }
  \label{fig: scenario2-r1r4}
\end{figure}

\textbf{The Goal of Experiment.}
In the original study of fit estimation~\cite{reimann_visual_2021}, strong decentering was described as the data shifting by 0.75 times the noise size, where points were still on both sides of the line as shown in the OSF materials. 
The results indicate that 1) the ratings for \textit{(strong decentering, 0.2 noise)} (R1) are higher than those for \textit{(no decentering, 0.4 noise)} (R2); and
2) the scores for \textit{(strong decentering, 0.4 noise)} (R3) are on par with those for \textit{(no decentering, 0.6 noise)} (R4).
We doubted the effect of decentering and plan to test a more extreme decentering (shift value $=$ 0.99) scenario (\cref{fig: scenario2-r1r4}).
It would make points almost clustering on one side of the line as shown in the R1 and R3 of \cref{fig: scenario2-r1r4}. 
As outlined in \cref{sec: fit}, agent ratings supported \textbf{C2.4} and revealed similarities to human judgments in specific local comparisons (\cref{sec: fit replication}). 
Thus, we aimed to compare the scores from agents for extreme decentering against real user study outcomes to investigate its alignment in swiftly adjusting experimental parameters.

\textbf{Experiment Design.}
We conducted a new within-subject study with a 3 (noise levels: 0.2, 0.4, 0.6) $\times$ 3 (decentering conditions) design, involving 38 participants.
One sample of suspicious rating pattern was removed for short completion time and casual ratings across all images.
The procedure and instructions replicated those in the original study, using an online interface integrated with questionnaires and visual stimuli. 
Each participant went through 9 scatterplots in a randomized order and rated via a slider bar.
This procedure ensures a balanced dataset for ANOVA and covers R1-R4 (\cref{fig: scenario2-r1r4}).
For the agent-based study, due to the input limit, we excluded the three images with \textit{(no decentering, 0.2 noise)} and \textit{(strong decentering, 0.6 noise)} from the agent's input, as they were irrelevant to R1-R4. 
We used GPT-4V to complete the within-subject ratings for six images with 20 repetitions to mitigate randomness.

\textbf{Results.}
We combined the ratings for upward and downward decentering (there was no effect of direction) into R1 and R3 by averaging, as the original study did. 
For the study involving human subjects, a repeated measures ANOVA analysis revealed a significant main effect of decentering, $F(1, 36) = 14.00, p < .001, \eta_{p}^{2} = .32$, and noise, $F(1.6, 57.60) = 97.047, p < .001, \eta_{p}^{2} = .73$ (both Greenhouse-Geisser corrected). 
No significant interaction effect was found, $F(1.90, 68.22) = .003, p = 1.00, \eta_{p}^{2} = .00$.
Graphical inspection indicated only minor deviations from the normality assumption required for ANOVA.
These conclusions are consistent with those presented in the original study.

\begin{figure}[t]
  \centering
  \includegraphics[width=\linewidth]{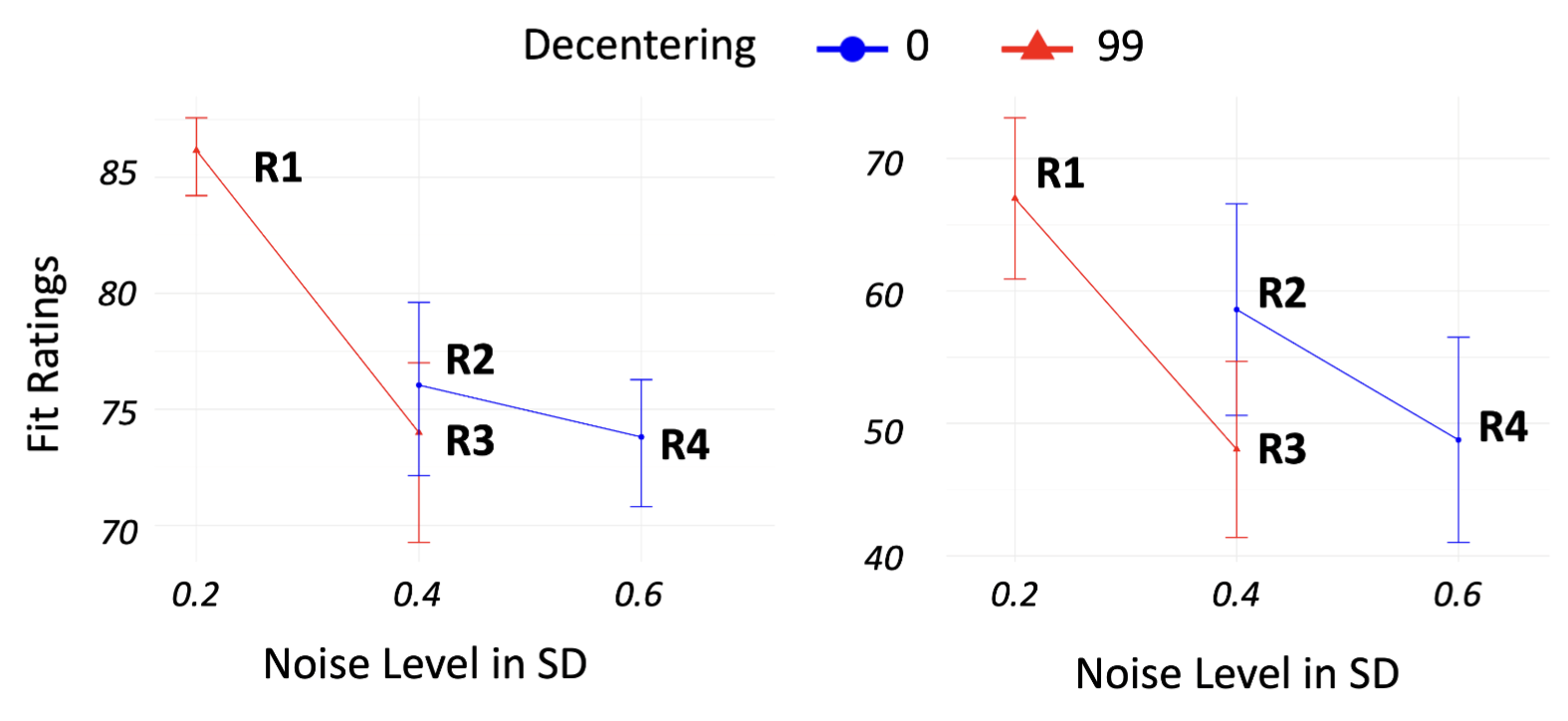}
  \caption{Fit ratings by agent (left) and human (right) for different conditions of noise and decentering. Error bars represent 95\% CIs. }
  \label{fig: scenario2-results}
  \vspace{-12px}
\end{figure}

\cref{fig: scenario2-results} shows the ratings of agent (left) and humans (right). 
The results of the new user study align with the original study. 
Even if decentering reaches a more extreme value of 0.99, the result of R1>R2>R3=R4 is still valid. 
Agent feedback is also consistent with the two user studies, and the impact of decentering on it is still weaker than that on humans.

\textbf{Implications.}
Our experiments showed two potential use cases. 
If agent outputs align with the existing user study, and new studies only adjust experimental parameters without altering other procedures, reflecting these parameter changes in the agent's input may allow for the preliminary simulation of the new studies.
In addition, if the results of a small-scale pilot study align with the predictions by the agent, evaluators may adjust parameters and watch for significant output changes to set parameters for the formal study.
This approach assists in setting parameters that align with experimental objectives.
However, more empirical evidence from studies with large datasets is needed to further confirm this finding. 

%% file: sections/7-discussion.tex
\section{Discussion}
\label{sec: discussion}
\textbf{Possible misuses.}
Novice evaluators may be influenced by the agent's human-like outputs as demonstrated in our study, despite their inherent unreliability. 
Moreover, verifying qualitative data proves more difficult than text data, possibly amplifying reliability concerns regarding crowdsourced data~\cite{hamalainen_evaluating_2023}.
Currently, only closed-source large models like GPT-4V can rate visualizations in a human-like manner, with high usage costs serving as a temporary barrier.
However, rapid advancements in open-source models may soon reduce these barriers, increasing associated risks.

\revised{
\textbf{Keeping Pace with State-of-the-Art LLMs.}
More advanced LLMs are emerging that could produce ratings more aligned with human responses. 
We reproduced the agent study in \cref{sec: fit replication} using GPT-4o and presented the results in the supplementary material. 
While GPT-4o differs from GPT-4V in rating range, its alignment with humans mirrors that of GPT-4V. 
We believe our study methods (\cref{sec: general modifications}) and open-source code can be directly applied to any LLM, and encourage others to use them to quickly verify the alignment of the latest agent ratings with human responses.

}

\textbf{Data pollution.}
Since our studies predate GPT-4V's latest training, data pollution was anticipated and became evident in our test on Imputation for Uncertainty using text and general descriptions (\cref{sec: extra test}).
However, the agents' poor performance with specific image inputs made this issue less pronounced.
Interestingly, data pollution offers an expected benefit. 
Though it hinders simulating new expert research, it aligns agent feedback with general evaluators for basic tasks, potentially making them useful for everyday visualization design and evaluation.

\label{sec: potential directions}
\textbf{Knowledge-assisted and reliable agent simulation.}
As discussed in \cref{sec: extra engineering}, injecting external knowledge can enhance the alignment of agent-based feedback in certain scenarios but raises concerns about generalizability and reliability. 
\revised{Content searched from websites could introduce additional biases}, especially for non-expert evaluators \revised{unable to identify them}.
The challenge stems from the scattered and unsystematic nature of existing knowledge on human perception, with no comprehensive visual cognitive knowledge readily available. 
We attempted to use ``Visualization Psychology''~\cite{szafir_visualization_2023} and its references as a knowledge source for RAG to extract relevant information automatically.
However, this failed due to semantic mismatches between visualization designs and knowledge, as well as uncertainty about relevant knowledge availability and selection.
Reliable agent feedback will require advances in visualization research to provide systematic, practical knowledge for evaluation.

\textbf{Profile-aware agents for diversity simulation.}
Simply modifying demographic data, such as age or gender, does not affect agent outputs (\cref{sec: pilot study-ratings}) since such data is not emphasized during models' training.
Due to privacy concerns, research often excludes detailed user profiles. To enable models to handle diverse user profiles, comprehensive first-hand data collection is necessary.
For example, gathering complete user study logs from college students and fine-tuning large models could allow agents to emulate different student profiles.
This method could be expanded to other groups, enabling profile-aware diversity in outputs and modeling intra- and inter-group variations.

\textbf{Effective input preprocessing and prompting strategies.}
Our further tests (\cref{sec: extra test}) show that factors like modifying prompt keywords and mentioning scenes significantly influence agent ratings.
\revised{Similar points have also been found in recent or concurrent work~\cite{wang_vis24_howaligned,li2024visualizationliteracy, lo_vis24_howgood} when exploring models' accuracy and chart takeaways.}
While we did not explore all factor combinations, future research should carefully consider preprocessing and prompt strategies.
Furthermore, because other open-source LMMs are less capable, making them unsuitable for thorough testing, we could not confirm the generalizability of our results. 
Future use of large model-based agents will require preliminary studies to gather targeted insights for subsequent optimization.

\textbf{Flexible agent-based procedure design.}
The design of our agent-based study was constrained by the current capabilities of agents (\cref{sec: general modifications}). However, the emergence of advanced agents with visual capabilities and memory retention across multiple dialogue rounds could expand design possibilities.
These agents could simulate a more human-like process by using memory to mimic how people rate images in a temporally continuous, between-subjects manner. 
Additionally, with scaling laws driving rapid improvements in large models and increasing context token capacities, future agents may overcome current limitations on the number of image inputs, enabling larger-scale within-subject study designs.

\revised{
\textbf{Simulating broader feedback in broader scenarios.}
We investigated agent-based study procedures for quantitative data, applicable to broad evaluations using Likert-scale rating.
While most OSF records lack user think-aloud data, some studies, like those on timelines~\cite{di_bartolomeo_evaluating_2020}, have documented participant strategies in natural language, warranting further investigation.
In addition, our study excludes interactive or complex decision-making tasks and leaves a gap in evaluating interactive systems.
For such systems, agents can be prompted to understand interfaces and tasks, thus enabling interaction sequences for task completion.
This approach can be integrated with computational models to assess time overhead or combined with our study to simulate overall ratings.
Systems requiring domain expertise and complex decision-making, such as visual analytics, demand agents equipped with domain-specific knowledge and advanced prompt chains~\cite{grunde-mclaughlin_designing_2023}.
}

\textbf{Limitation and future work.}
Our study is limited to six existing studies due to the restricted availability of open resources.
Additionally, although our incremental experiments complemented each other, they did not offer a complete picture.
Specifically, we did not delve deeply into qualitative analysis of textual content, which could provide richer insights.
We advocate for future empirical studies to include detailed experimental records to facilitate research on automated agent-based methods.
we aim to utilize larger datasets and advanced agent techniques to keep pace with rapid advancements in the field.

%% file: sections/8-conclusion.tex
\section{Conclusion and Takeways}
\label{sec: takeaways}

Based on all the findings from the study, we answer the 3 \textbf{RQ}s as:
\begin{itemize}[leftmargin=10pt]
    \item \textbf{RQ1 (What are the characteristics of agent ratings?):
    Agents can simulate ratings relatively aligned with human responses.} Large model-based agents are now aggregations of experiences, relying on a general understanding of visual graphs to mimic human outputs. Their ratings show promise in aligning with human ones, though they cannot yet capture profile diversity (\cref{sec: pilot study}).
    \item \textbf{RQ2 (How well do agent and human ratings align): The alignment of agent ratings is positively related to the confidence level of human experts for giving hypotheses.} Agent ratings are more likely to align only when our consulted external expert evaluators can offer consistent hypotheses with high confidence for the experiment (\cref{sec: rq2}).
    \item \textbf{RQ3 (How to better align agent and human ratings): Constructive strategies applied to the agent inputs can influence alignment.} Beyond directly utilizing materials in user study as the agent's textual and visual inputs, preprocessing them or integrating external knowledge could change rating alignment. Still, more research is needed for validation (\cref{sec: rq3}).
\end{itemize}

Given the previous insights and the potential scenarios we have studied, we identify three more key takeaways:
\begin{itemize}[leftmargin=10pt]
    \item \textbf{Potential Scenario.} Using large model-based agents in facilitating fast, iterative design of experiments is promising when their ratings can align with those of prototype experiments (\cref{sec: potential scenarios}).
    \item \textbf{Be Cautious.} 
    According to current results of current techniques of large models, agents cannot process visual details like humans through vision or simulate diverse users (\cref{sec: pilot study}).
    They cannot simulate most conclusions in visualization research since experts often lack confidence in their hypotheses and agent ratings are less likely to align with human data (\cref{sec: rq2}).
    We also need to cautiously handle adjustments to input and pipeline, mindful of their impact on alignment. (\cref{sec: rq3}).
    Regardless, agent ratings require validation with real user data.
    \item \textbf{Promising Future.} Leveraging their strong reasoning, future advancements in large models' visual and cognitive functions promise wider applications for more aligned ratings and feedback across more scenarios (\cref{sec: potential directions}).
\end{itemize}

%% file: meta/appendix.tex
\setcounter{section}{0}
\section*{Appendix}
\label{sec:appendix}
\pagestyle{empty}

\definecolor{backgroundcolor}{RGB}{240, 243, 248} %
\definecolor{bracescolor}{RGB}{255, 127, 80} %
\newcounter{Prompt}[section]
\newenvironment{Prompt}[1][]{%
  \refstepcounter{Prompt}%
  \ifstrempty{#1}{%
    \mdfsetup{%
      frametitle={%
        \tikz[baseline=(current bounding box.east),outer sep=0pt]%
        \node[anchor=east,rectangle,fill=blue!20] {\strut \color{DarkOliveGreen}{Prompt}~\thePrompt;};%
      }%
    }%
  }{%
    \mdfsetup{%
      frametitle={%
        \tikz[baseline=(current bounding box.east),outer sep=0pt]%
        \node[anchor=east,rectangle,fill=blue!20] {\strut \color{DarkOliveGreen}{Prompt}~\thePrompt:~\color{Maroon}{#1}};%
      }%
    }%
  }%
  \mdfsetup{%
    innertopmargin=3pt,
    linecolor=blue!20,
    linewidth=1pt,
    topline=true,
    frametitleaboveskip=\dimexpr-\ht\strutbox\relax,
    backgroundcolor=backgroundcolor %
  }%
  \begin{mdframed}[]\relax%
    \lstset{%
      literate={\{}{{\textcolor{bracescolor}{\{}}}1 %
               {\}}{{\textcolor{bracescolor}{\}}}}1,
      basicstyle=\ttfamily,
      breaklines=true,
      backgroundcolor=\color{backgroundcolor}, %
    }%
}{%
  \end{mdframed}%
}

\newcommand{\highlightbrace}[1]{\textcolor{bracescolor}{#1}}
\newcommand{\highlighttext}[1]{\textcolor{blue}{#1}}

In the first section, we present the prompt and codes for two experiments (Fit Estimation and Imputation for Uncertainty) as references. In the second section, we present the details of other experiments analysis.

\section{Prompts and Code}

We sequentially showcase the prompts and code utilized in these experiments.
Within the prompts, "ROLE" refers to the desired background role for the large model, "TASK" denotes the specific task for user in experiment, "RATING" outlines the criteria for rating, and "OUTPUT FORMAT" specifies the desired output format.
ROLE is summarized based on descriptions of participants in the paper, while TASK and RATING are generated according to descriptions in the original text and tutorials. OUTPUT FORMAT is created based on our own data requirements.
In some experiments (e.g., Fit Estimation), the specific task is rating, thereby combining TASK and RATING into a single prompt.

The code primarily displays the handling of raw data, the organization of prompts, and the workflow of the experiments.
In the Fit Estimation experiment, as the specific task itself involves rating, the experimental code comprises only one operation. Conversely, in the Imputation for Uncertainty experiment, since the specific task and the rating task are distinct, we first instruct the large model to outline the specific steps necessary to complete the task in the first operation, followed by rating these steps in the second operation.

\subsection{Fit Estimation}
\vspace{1em}
\begin{Prompt}[ROLE]\label{role}
You are an average user.
\end{Prompt}

\vspace{0.5em}

\begin{Prompt}[TASK(RATING)]\label{task}
Here are \highlightbrace{\{}\highlighttext{number}\highlightbrace{\}} scatterplots. You are instructed to estimate how well the data points fitted to the shown line. Your score ranges from 0 (very low fit) to 100 (very high fit), a perfect fit would be to have all data points on the line. Please list the scores and the reason in the order of the images.
\end{Prompt}

\vspace{0.5em}

\begin{Prompt}[OUTPUT FORMAT]\label{output format}
Please give an additional scoring result in json format at the end of your answe, like json[score1, score2, ...]. 
\end{Prompt}

\begin{lstlisting}[language=Python]
name = "fit_estimation_experiment1"
file_path = "./fit/stimuli_exp1+exp2/"
sd_list = ["exp1_sd05", "exp1_sd10","exp1_sd15","exp1_sd20","exp1_sd25",]
index_list = ["1", "2", "3", "4", "5", "6"]
for i in index_list:
    description = "parallel_experiment_" + i
    image_urls = [file_path + sd + "_" + i + ".png" for sd in sd_list]
    prompt = ROLE + \
        TASK.format(number=len(image_urls)) + \
        JSON_FORMAT
    # Disrupting the order of images and calling visual models
    run_test(
        name=name,
        description=description,
        prompt=prompt,
        image_urls=image_urls,
        logger=logger,
        times=5)
\end{lstlisting}

\subsection{Imputation for Uncertainty}
\vspace{1em}
\begin{Prompt}[ROLE]\label{role}
You are an average user.
\end{Prompt}
\vspace{0.5em}
\begin{Prompt}[TASK]\label{task}
The graphs show the relationship between two variables X and Y. The points belong to two groups Cyan and Orange.   
However, the original dataset has certain missing values in either the X or Y for every variable. We use a statistical method called imputation to estimate these points. The point represents the most likely (average) estimate of the missing data point derived from the imputation method. However, as imputation cannot precisely estimate values for the missing data points, we also get some uncertainty from our imputation method.
There are five different uncertainty representations, each for a graph provided.

In the first graph \highlightbrace{\{}\highlighttext{first}\highlightbrace{\}}.

In the second graph \highlightbrace{\{}\highlighttext{second}\highlightbrace{\}}.

In the third graph \highlightbrace{\{}\highlighttext{third}\highlightbrace{\}}.

In the fourth graph \highlightbrace{\{}\highlighttext{fourth}\highlightbrace{\}}.

In the fifth graph \highlightbrace{\{}\highlighttext{fifth}\highlightbrace{\}}.

With the introduction above, you need to understand the visualizations and finish the tasks based on the visualization.

Tasks: For each uncertainty representation, complete the following task:
\highlightbrace{\{}\highlighttext{task}\highlightbrace{\}}
You need to tell me the exact process of completing the task using the visualizations and the difficulties you may encounter, and DO NOT give your final answer of the task.
\end{Prompt}
\vspace{0.5em}
\begin{Prompt}[RATING]\label{rating}
Based on the background, someone gives the following analysis: \highlightbrace{\{}\highlighttext{steps}\highlightbrace{\}}. From the human perspective you are simulating, for each of the representations, rate your confidence in your answer on a scale of 1 to 5 with 5 representing the highest confidence. As an average human participant, DO NOT state individual variance on this task, and you MUST give an specific answer.

\end{Prompt}
\vspace{0.5em}
\begin{Prompt}[OUTPUT FORMAT]\label{output format}
Please give an additional result in json format at the end of your answer, like 
json[\{"steps": steps, "difficulties": difficulties (also provide your reason inside this value)\},
    \{"steps": steps, "difficulties": difficulties (also provide your reason inside this value)\},
    ...
].

\end{Prompt}

\begin{lstlisting}[language=Python]
# Data Preparation
tasks = {"mean": MEAN_TASK, "trend": TREND_TASK}
task_description = {"mean": "Based on the understanding of how missing data is imputed, estimate the average value of X of the whole dataset.", "trend": "Based on the understanding of how missing data is imputed, identify the trend line which best represents the relationship between x and y."}
type_list = ["none", "mean", "ci", "density", "gradient"]
proportion_list = ["30", "50"]
error_list = ["15", "20"]
mean_number_list = ["graph-x-001", "graph-y-002"]
trend_number_list = ["graph-xy-001", "graph-xy-002"]
file_path = "./uncertainty/experiment_preparation/images/"
image_description = [
    " (none), data points with missing values are not shown as they cannot be accurately plotted on the chart",
    " (mean), the missing points are estimated, and the mean of these estiamtes are represented as hollow points",
    " (ci), we represent this uncertainty using the bars which represent the 95%
    " (density), we represent this uncertainty using the probability density plots",
    " (gradient), we represent this uncertainty using the gradient plots showing 95%
]
\end{lstlisting}
\begin{lstlisting}[language=Python]
# Execution
name = "uncertainty_experiment"
task = "mean"
for proportion in proportion_list:
    for error in error_list:
        for number in mean_number_list:
            description = f"{task}_proportion{proportion}_error{error}_{number}_step1"
            image_urls = [f"{file_path}{_type}/{task}/proportion{proportion}/error{error}/{number}.png" for _type in type_list]
            prompt = ROLE + \
                    tasks[task] + \
                    JSON_FORMAT
            responses = run_test1(
                name=name,
                description=description,
                prompt=prompt,
                image_urls=image_urls,
                image_description=image_description,
                logger=logger1,
                times=5)
            
            for response in responses:
                prompt = ROLE + \
                    tasks[task] + \
                    RATING.format(steps=response) + \
                    RATING_JSON_FORMAT
                description = f"{task}_proportion{proportion}_error{error}_{number}_step2"
                run_test2(
                    name=name,
                    description=description,
                    prompt=prompt,
                    image_urls=image_urls,
                    logger=logger2,
                    times=1)
\end{lstlisting}

\begin{lstlisting}[language=Python]
# Execution
name = "uncertainty_experiment"
task = "mean"
for proportion in proportion_list:
    for error in error_list:
        for number in mean_number_list:
            description = f"{task}_proportion{proportion}_error{error}_{number}_step1"
            image_urls = [f"{file_path}{_type}/{task}/proportion{proportion}/error{error}/{number}.png" for _type in type_list]
            prompt = ROLE + \
                    tasks[task] + \
                    JSON_FORMAT
            responses = run_test1(
                name=name,
                description=description,
                prompt=prompt,
                image_urls=image_urls,
                image_description=image_description,
                logger=logger1,
                times=5)
            
            for response in responses:
                prompt = ROLE + \
                    tasks[task] + \
                    RATING.format(steps=response) + \
                    RATING_JSON_FORMAT
                description = f"{task}_proportion{proportion}_error{error}_{number}_step2"
                run_test2(
                    name=name,
                    description=description,
                    prompt=prompt,
                    image_urls=image_urls,
                    logger=logger2,
                    times=1)
\end{lstlisting}

\section{Other Experiments Analysis}
In this work, we conducted experiments on a total of six papers. In the main body of the paper, we present the detailed analysis process of three experiments along with some additional experiments. The remaining three experiments and further additional experiments are displayed as follows.
Besides, we also conducted one replication experiment on Fit Estimation with GPT-4o, where we show the results at last.

\subsection{Experiment: Timeline}
\subsubsection{Original Study}
\textbf{Material and Procedure.} Bartolomeo et al.~\cite{di_bartolomeo_evaluating_2020} assessed the effect of different timeline shapes on participants' task performance.
They considered four different timeline shapes (horizontal line, vertical line, spiral, and circle) and formed three types of temporal event sequence data (recurrent, non-recurrent, mixed) and four tasks (when, what, find, and compare).
Each participant was assigned to a certain task type and completed the tasks regarding all types of timeline shapes and events.
Participants' completion time, correctness, and easiest-to-read selection were collected as outcomes.

\textbf{Results.}
Concerning readability, their expectations are structured around the types of visualization, data, and tasks, across three levels: 
1) horizontal timelines are preferred by users (\textbf{H4.1});
2) In terms of data type impact, linear timelines are more readable for linear data, whereas circular and spiral timelines are seen as more suitable for recurrent and mixed datasets (\textbf{H4.2}); 
3) No specific assumptions are made regarding individual tasks. 
Participants' easiest-to-read selection is regarded as qualitative results, the distribution of which is directly analyzed in the paper. 
Empirical evidence illustrates that \textbf{H4.1} was supported as \textbf{C4.1}, while \textbf{H4.2} was not supported as the horizontal line was generally more perceived and the recurrent dataset only showed mixed responses \textbf{C4.2}. 
Besides, the mixed responses depending on the tasks didn't show any definitive claim (\textbf{C4.3}).

\subsubsection{Agent-based Study}
\textbf{Material and Procedure.} Given that the original readability assessment involved presenting four timelines simultaneously, we were able to replicate this process exactly. 
For each trial, we prompt agents to begin with ``You are an average user'' for role-play and then input the tutorial description and a question along with the four visual timelines into the agent and repeat this ten times.

\textbf{Results.}
The results indicate that the agent predominantly selected horizontal timelines, doing so with a 96.97\% probability, while vertical lines were chosen in the remaining 3.03\% of cases (all in the ``find'' task with ``mixed'' dataset).
This aligns with \textbf{H4.1} and support \textbf{C4.1}.
Although this supports \textbf{C4.2}, the agent's results are significantly different from humans, as spiral or circle timelines have never been selected.

\subsection{Experiment: Texture}
\subsubsection{Original Study}
\textbf{Material and Procedure.}
He et al.~\cite{he_design_2024} studied design characterization for textures in visualization.
The first study was to develop the designs for later experiments, which is beyond our work.
Their second study assessed the ratings of the visual appeal of the designs. 
They employed a between-subjects method on chart type (bar, pie, map) and a within-subjects method on texture type (geometric, iconic).
Participants were randomly assigned to one chart type and completed two blocks of evaluation, each block presenting one texture type and consisting of four images. 
Specifically, they rated the vibratory effect (which is briefly explained to them in the study) and five items for Beauvis~\cite{he_beauvis_2023} and also ranked the designs.
The third study concerned the readability of the designs.
Participants sequentially answered the question about ``more'' or ``fewer'' and rated each fill type (unicolor, iconic, geometric) $\times$ chart type (bar, pie charts) condition on readability, as well as the aesthetic ratings after reading.

\textbf{Results.}
In the second study of visual appeal, the distributions of the collected ratings and rankings are reported, as well as their means and bootstrap CIs.
The ratings were primarily used for selecting visual stimuli and no related hypothesis was set.
Their findings include iconic maps that received the lowest ratings (\textbf{C5.1}) and relations between ratings and word-scale distribution (\textbf{C5.2}).
For the third study, they performed pairwise comparisons with bootstrap CIs.
They hypothesized that both iconic and geometric textures serve as more readable than unicolor (\textbf{H5.3}) and iconic textures would be perceived as more aesthetically pleasing (\textbf{H5.4}).
Both hypotheses were not fully supported.
For bar charts, unicolor textures significantly improved readability, whereas, in the case of pie charts, the opposite was true (\textbf{C5.3}).
Additionally, for bar charts, geometric textures were viewed as less appealing than iconic ones, with no marked difference for pie charts (\textbf{C5.4}).

\subsubsection{Agent-based Study}
\textbf{Material and Procedure.}
Since the original experiments were originally a within-subjects experiment with four pictures in a batch, we were able to completely replicate the human process and let the agent repeat each group ten times. 
We prompted the agents with additional role-play instructions and background information about BeauVis and the vibratory effect, in addition to the original stimuli from OSF.

\textbf{Results.}
In the visual appeal experiment, the results show that the distribution of ratings is not similar to that of the user study, and it does not confirm \textbf{C5.1} and \textbf{C5.2}. 
This shows the agent's missing visual skill for specific design details and not understanding map visualization well.
Also, we figured out the variance for Beauvis's five scores in every group. 
The biggest variance for all groups was under 0.3, showing the agent doesn't think about the differences between the five items like a human would.
For the readability experiment, The results partially align with \textbf{C5.3} on readability scores.
Unicolor has a significant advantage over both textures in readability scores on both charts.
The results of after-reading Beauvis' ratings do not support \textbf{C5.4}. 
In the bar chart, the aesthetic score for iconic textures is the lowest among the three and significantly lower than geometric ones, which contrasts to \textbf{C5.4}.
The results from agent-based experiments deviate from both evaluators' hypotheses and user study findings, underscoring the agent's limited grasp of aesthetic assessment and texture.

\subsection{Experiment: Magnitude Judgement}
\label{sec: axis manipulation}
\subsubsection{Original Study}
\textbf{Material and Procedure.}
Bradley et al.~\cite{bradley_magnitude_2024} manipulated the axis limits surrounding plotted data to learn about magnitude judgments.
In the first experiment, they created some pairs of scatter plots. Each pair presented data points of the same value, but the points were placed at a high or low physical position by changing the limits of the Y-axis.
The data corresponded to real-world scenarios for participants to rate the magnitude of them.
They employed a within-subjects design, where each participant encountered a series of scatter plot pairs and plots in one pair were arranged to be far from each other in case the participants could infer the intent of the study.
Participants rated the magnitude they perceived in a 7-point Likert scale.
In the second experiment, they tested scenarios in plots with inverted axis orientations to figure out whether physical position or relative position to the end of the axis has a pivotal effect on the participants' magnitude judgment. The experiment settings were the same as those in the first experiment except for the axis orientation. 

\textbf{Results.}
They constructed cumulative link mixed-effects models for the analysis of ratings.
In the first experiment, participants rated data of greater magnitudes higher when placed near the chart's top (\textbf{C6.1}). 
This aligns with established conceptual metaphors for magnitude and prior research that associates higher positions with higher values, which could be viewed as a hypothesis (\textbf{H6.1}).
In the second experiment, they observed the reverse ratings against the first study, indicating relative positions were the key factors rather than physical positions (\textbf{C6.2}).
The second experiment was conducted without a predefined hypothesis.

\subsubsection{Agent-based Study}
\textbf{Material and Procedure.}
Contrary to the original study, which aimed to minimize the chance of similar scenario versions appearing consecutively, our approach adopted a within-subject design (Sec~\ref{sec: general modifications}). 
This involved inputting two images from the same scenario into agents for rating each time. 
We prompted agents to begin with ``You are an average user'' for role-play and then inputted the original stimuli from OSF and repeated five times. 

\textbf{Results.}
Our agent-based procedure eliminates variables like participant and visualization literacy, and hence the original statistical model couldn't be applied. 
Instead, we mitigate scenario influence by comparing the high-to-low ratio within individual scenarios.
Unexpectedly, in the first experiment, agents consistently rated charts higher when positioned high, contrary to both \textbf{H6.1} and \textbf{C6.1}, offering an anti-human interpretation of axis comparisons. 
In the second experiment, with inverted axes, agents maintained this perspective 90\% of the time, again opposing \textbf{C6.2}.
Agent's focus on axis scale text prompted us to pay closer attention to textual elements in images in future experiments.

\subsection{Additional Experiment: Timeline}

\textbf{Explicit comparisons of variables may lead the agent to rely on the text to offer stereotypes similar to those of experts.}
When stating the current dataset type without requiring comparison with other scenarios, the conclusion that horizontal timelines are best remained unchanged (92.42\%).
Furthermore, only when the prompt explicitly stated the different ratings for comparing three types of datasets did the agent consider spiral/circle timelines better for recurrent datasets; 
for the other two types, vertical and horizontal were deemed equally good, without distinction, as illustrated by the agents' explanation. 
These phenomena align with the evaluators' original hypothesis \textbf{H4.2} and contradict the findings \textbf{C4.2}.
Our observations indicate that this outcome is independent of whether images are provided, suggesting that explicit comparisons of variables may lead the agent to rely on the text to offer stereotypes similar to those of experts.

\subsection{Further Experiment with GPT-4o}

\begin{figure}[ht]
  \centering
  \includegraphics[width=\linewidth]{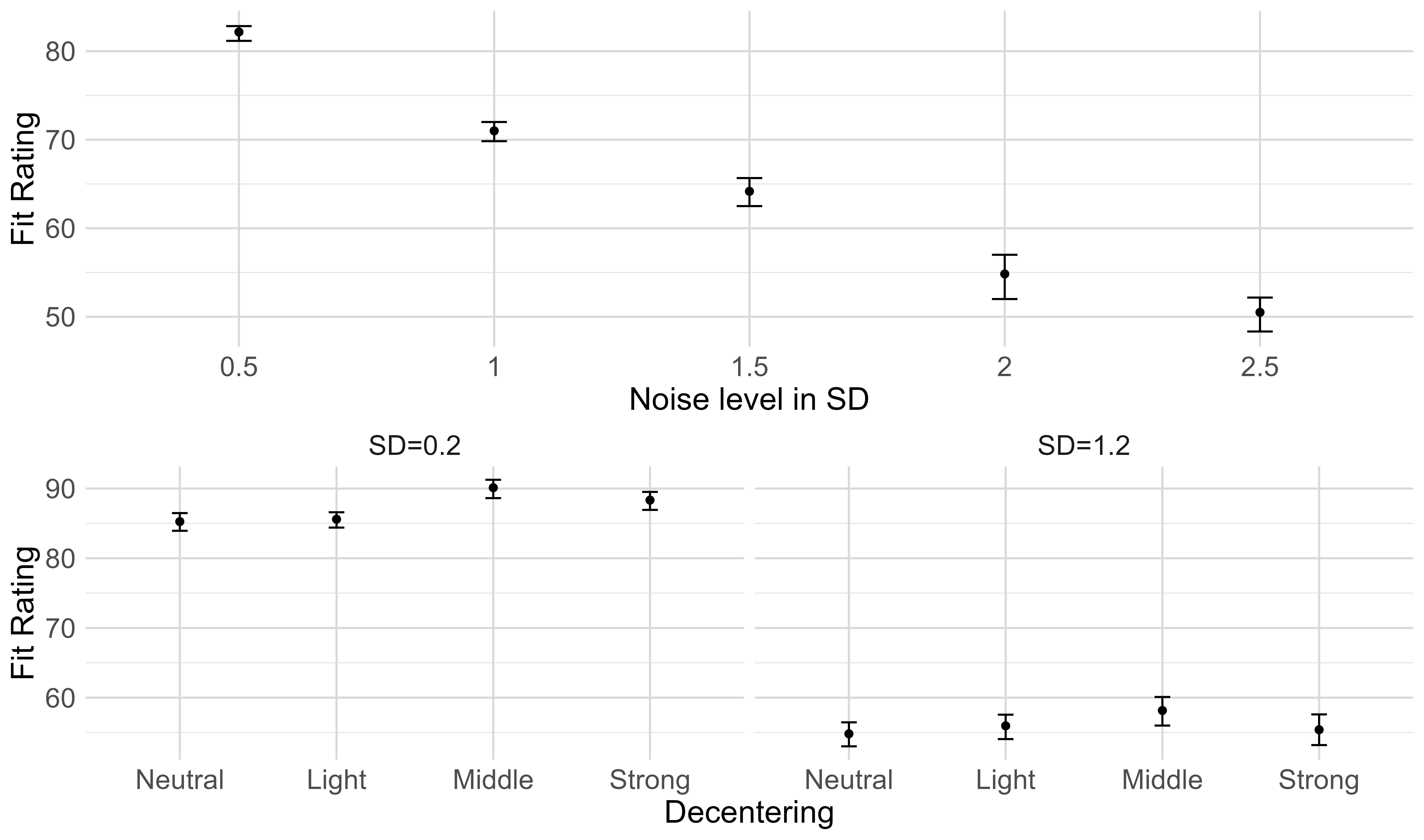}
  \caption{Results of GPT-4o's ratings for fit estimations under various noise levels (top) and decentering (bottom). Error bars represent 95\% bootstrap CIs.}
  \label{fig: fit-gpt4o}
\end{figure}

Comparing Fig.~\ref{fig: fit-gpt4o} and Fig.~\ref{fig: fit} in the main paper, for different noise levels, GPT-4o’s ratings are more consistent across sessions for the same condition, with smaller CIs, and show larger differences under varying conditions compared to GPT-4v’s ratings. 
This trend also holds across different decentering conditions. 
Additionally, GPT-4o’s ratings exhibit a relationship with noise levels that is closer to a negatively accelerated pattern. 
Overall, however, the alignment of GPT-4o with human conclusions in the Fit Estimation experiment is similar to that of GPT-4v.